\documentstyle[prl,aps,multicol,epsfig]{revtex}
\begin{document}
\draft
\preprint{HD-THEP-00-39}
\title{Phase Transitions in liquid Helium 3}
\author{Markus Kindermann \cite{mailkind} and Christof Wetterich\cite{mailwett}}
\address{
Institut f\"ur Theoretische Physik,
Universit\"at Heidelberg,
Philosophenweg 16, 69120 Heidelberg,
Germany
}
\date{\today}
\maketitle

\begin{abstract}
The phase transitions of liquid $^3He$ are described by truncations of an exact nonperturbative renormalization group equation. The location of the first order transition lines and the jump in the order parameter are computed quantitatively. At the triple point we find  indications for partially universal behaviour. We suggest experiments that could help to determine the effective interactions between fermion pairs.
\end{abstract}

\begin{multicols}{2}

\narrowtext

The low temperature phase transitions and the superfluid phases of liquid Helium 3 \cite{Osheroff} can be described \cite{Love} by a field theory for complex  $3 \times 3-$ matrices $A$ representing fermion pairs. An approximation based on quartic polynomials in $A$, including the renormalization group running of the coupling constants \cite{Sokolov79}, can account for the rough phase structure but not for details of the transitions. We will be concerned with the transitions from the normal liquid to superfluidity in vanishing external magnetic field. In a mean-field treatment those transitions are of second order, whereas early RG-calculations \cite{Sokolov79}  found hints for them to be fluctuation induced first order transitions. The BCS weak-coupling theory estimates the critical region where fluctuations invalidate the mean-field approximation to extend over a temperature interval $ \triangle T \approx 10^{-8} K$. Based on experiments on zero sound absorption \cite{Wheatley} it has been conjectured, however, that the critical region might be up to a thousand times larger \cite{Sokolov83}.
The theoretical answers therefore hinge on two problems: first, the ``microphysical'' effective interactions at the scale of the fermion pairs ($\Lambda^{-1} \approx 120  \AA $) are poorly known and second, the transition from the microphysical interactions to the macrophysics of the phase transitions (i.e. the thermodynamic potentials) is very complex. Only if the second problem can be solved, the experimental observations of the macrophysics (phase structure, jump in the order parameter, etc.) can be used to constrain the microphysical interactions. In this letter we propose a quantitatively reliable mapping of the microphysics onto the properties of the free energy density. This is effected by means of extended truncations to an exact nonperturbative renormalization group equation \cite{Wetterich}, in contrast to largely uncontrolled approximations used earlier \cite{Sokolov83,Bailin}. 

The different phases of $^3He$ are characterized by expectation values $<\!\! A\!\!>$ (order parameters) in different directions: the BW-state, $A_1=\boldmath{1}$, the ABM-state, $A_2=\sqrt{\frac{3}{8}}\;(\lambda_7+ i\lambda_6+\lambda_4 - i\lambda_5)$, and the planar state, $A_3=\sqrt{\frac{3}{2}}\;i\;\lambda_2$ represent the phases of the mean-field diagram. The $\lambda_i$ are the Gell-Mann matrices. Additionally, we examine the direction    $ A_4=\sqrt{\frac{3}{4}}\;(\lambda_3-i \lambda_1)  $.
Here, $< \!\! A \!\!>$ corresponds to the minimum of the effective potential $U(A)$ for spatially homogeneous fields. Our main aim is therefore the computation of $U(A)$ for a given microscopic effective action (at the scale $\Lambda$) for which we assume the conventional form \cite{Love}

\begin{equation}
 \Gamma_{\Lambda} =  \int{d^3 x \;\; \{ \;Tr \; \vec{\nabla} A^\dag \vec{\nabla} A + m \rho + \sum_{i=0}^4{b_i\;  I_i}  + a K_D \} } \label{eq:start}
\end{equation}
with $\rho=Tr\;A^\dag A$ and  $I_i$ the invariants to fourth order allowed by the symmetry group $G=SO(3)\times SO(3) \times U(1)$:

$ I_0  =  \rho ^2  $,
$ I_1  =  |Tr\; A^T A|^2  $,
$ I_2  =  Tr \;(A^T A)(A^T A)^*  $,
$ I_3  =  Tr \; (A^\dagger A)^2  $ and
$ I_4  =  Tr \; (A^\dagger A)(A^\dagger A)^* $.

We include the dipole interaction 

 \begin{equation}
 K_D=Tr \;A^\dagger \; Tr\; A + Tr\;(A^* A) 
\end{equation}
as a symmetry breaking perturbation arising from the spin-orbit coupling of the atoms in a Cooper pair and neglect the strain gradient terms \cite{Love}. All quantities are in units of a characteristic critical temperature $T_c=2.6\; mK$ and  $m(T) = m(T_c) + \frac{T-T_c}{T_c} + ... $  reflects the temperature dependence. In the paramagnon theory the quartic couplings $b_i$ are given by \cite{Sokolov79}

$$
b_0 = c(2+.2 \delta),\;\;\;
b_1 = -c(1+.1 \delta),\;\;\; 
b_2 = c(2-.05 \delta),\;\;\;
$$
\begin{equation}
b_3 = c(2-.55 \delta),\;\;\;
b_4 = -c(2+.7 \delta) 
\end{equation}
with $c=0.001$ and $\delta$ parametrizing the pressure dependence. We study a pressure regime of $-2<\delta<1.8$. For $\delta>1.9$ the microscopic potential becomes unbounded from below. This will be compared to a calculation where the couplings are ten times as strong.

Our approach is based on the effective average action $\Gamma_k[A(x)]$ which interpolates between the microphysics ($k=\Lambda$) and the macrophysics ($k=0$) by means of an exact flow equation \cite{Wetterich}. We truncate the most general functional dependence on $A(x)$ by

\begin{equation}
 \Gamma_{k} =  \int{d^3 x \;\;\;\; \{ Z_k\;Tr \; \vec{\nabla} A^\dag \vec{\nabla} A \;\; + \; U_k(A) + a_k K_D \} } \label{eq:trunc}
\end{equation}
where $U_k(A)$ is a function of the eighteen real fields $\varphi_y$ in $A$  respecting the  symmetry group $G$. We can compute the effective potential $U(A) \equiv U_0(A)$ by a numerical integration of the flow equation  \cite{Wetterich}

 \begin{equation}
  \partial_t U_k(A) = \frac{1}{2} \int{\frac{d^3 q} {{2 \pi}^3} \;\;\partial_t R_k \;\;Tr \;((Z_k q^2+R_k)  + M^2_k(A))^{-1}}
 \label{eq:dtU}
 \end{equation}
Here, $t=\ln\;k$. The infrared cut-off

   \begin{equation} 
R_k(q)=Z_k q^2(1-e^{\frac{q^2}{k^2}})^{-1} 
\end{equation}
makes the flow equation ultraviolet and infrared finite. We need the mass matrix of second derivatives of $U$, $M^2_{y,z}= \frac{\partial^2 U}{\partial \varphi_y \partial\varphi_z}$. The flow equation for the wave function renormalization  $Z_k$ can be found in \cite{Tetradis}.

In a first investigation we have used a sixth order polynomial approximation for $U_k$ and searched for minima in all the fifteen directions with different residual symmetries given in \cite{Vollhardt}. In this scheme we found a phase transition into the direction $A_4$ for large $\delta$. We have convinced ourselves, however, that polynomial approximations do not allow reliable statements about the phase diagram. This is why we deal with an arbitrary dependence of $U_k$ on the field in a given direction $A= \xi A_n$ for which we write $U_k(\xi A_n)=V^{(n)}_k(\rho)$ with $\rho=3 \xi^2$. We assume, that the directions $n$ include  the absolute minimum of $U_k$.  We get the flow equation for $V^{(n)}_k$ by evaluating (\ref{eq:dtU}) at points on the line $\xi A_n$. To do so, we need to know the second derivatives of $U_k$ with respect to all fields, however. That is why we include in our truncation four coupling functions $b^{(n)}_i(\rho)$ multiplying all fourth order invariants other than $\rho^2$:

\begin{equation}
U(\xi A_n+\epsilon)=\{V^{(n)}(\rho) + \sum_{i=1}^4{b^{(n)}_i(\rho)(I_i - f^{(n)}_i \rho^2)}\} |_{\xi A_n+\epsilon} \label{eq:trunc1}
\end{equation}
The numbers $f^{(n)}_i $ are chosen such that the invariants $ I_i - f^{(n)}_i \rho^2$ do not contribute on the line $\xi A_n$.
Here  $\epsilon$ denotes fields orthogonal to $\xi A_n$ and we expect that this truncation  gives a good approximation to the true potential in a neighborhood of the line $\xi A_n$.

The flow equations of the coupling functions  $b^{(n)}_i(\rho)$ are derived by forming appropriate linear combinations of second and third derivatives of (\ref{eq:dtU}) with respect to certain fields. At the origin we need to take fourth derivatives to extract the couplings from our potential truncation. We have computed these flow equations for the four field directions $A_n$. The resulting equations are very long, altogether about 4000 lines. They all derive, however,  from  compact expressions, namely field derivatives of (\ref{eq:dtU}), which can be evaluated by a computer. We consider this structural simplicity to be one of the major strengths of our formalism.

 We have integrated the resulting equations numerically by laying the potentials and coupling functions on grids of ten points each. In a first run, we detected truncation errors in the quantities $\frac{\partial V^{(n)}}{\partial \rho}|_{A=0}$. They originate in an insufficient treatment of the symmetry constraints. We fight this problem in two steps. First, we infer the first potential derivatives at the origin from the term $m \rho$ and integrate a separate flow equation for $m$. Second, for the higher orders we employ for small $|t|$ a mixed truncation of the form

$$
 U(\xi A_n\;+\;\epsilon)=\{ m \rho + \sum_{i=0}^4{b_i \;I_i} + \sum_{j=0}^{10}{ c_j J_j} + \rho^4 w^{(n)}(\rho)
$$
\begin{equation}
 + \sum_{i=1}^4 {\rho^2 B^{(n)}_i(\rho) (I_i - f^{(n)}_i \rho^2)} \} |_{\xi A_n+\epsilon} \label{eq:trunc2}
\end{equation}
Here $J_j$ are the eleven invariants of sixth order. $w^{(n)}$ and $B^{(n)}_i$ describe the deviation from a sixth order polynomial approximation. The symmetry is manifestly respected up to sixth order. We have estimated the accuracy of the sixth order polynomial approximation by calculating $\partial_t U(\xi A_n)$ to fourth and sixth order in $\xi$ in two ways: from the flow of the couplings in the polynomial part of truncation (\ref{eq:trunc2}) and directly by equation (\ref{eq:dtU}). The discrepancy stays smaller than $20 \%$ as long as the flow parameter $|t|$ is not too big, $t>-5$. It grows to a few hundred percent, however, until the phase transition occurs at $t \approx -8$. This is due to terms of eighth and higher order that are truncated in the flow equations for the couplings $b_i$ and $c_j$ in (\ref{eq:trunc2}). We conclude, that with a polynomial truncation of the free energy to realistic order, no conclusive statements can be obtained about the phase diagram. As soon as the errors exceed $20\%$, we switch to the truncation (\ref{eq:trunc1}). The truncation error in the quantities $\partial_t V^{(n)}$ at the potential minima is estimated by comparing truncation (\ref{eq:trunc1}) with a truncation in which we extract the orthogonal masses from the polynomial truncation in the origin. The errors induced by the inaccuracies of $\frac{\partial V^{(n)}}{\partial \rho}|_{A=0}$ are of the order of magnitude of the truncation errors and both are not bigger than a few percent. This truncation should therefore allow us to compute the phase structure reliably.

\begin{figure}
\centering\epsfig{file=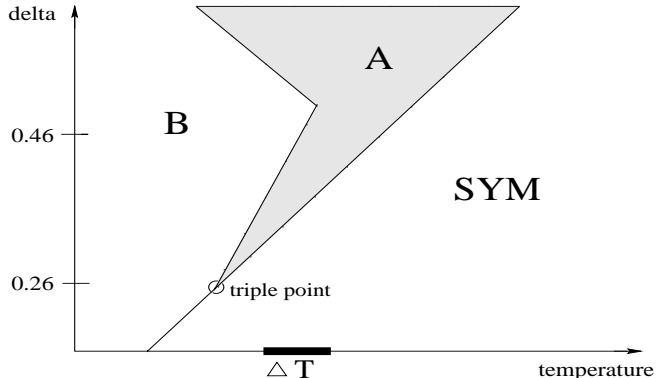,clip=,height=5cm,width=\linewidth}
\caption{Phase diagram of the effective theory for liquid $^3 He$ with strong coupling ($c=0.01$). The temperature scale is set by $\triangle T= 10^{-6} K$. The same picture is found for $c=0.001$ neglecting the spin-orbit interaction. Then $\triangle T = 10^{-8}$. }
\label{abb:PD1}
\end{figure}

First, we study the case of large couplings, $c=0.01$.  We neglect the spin-orbit coupling, which is justified in this case, the other masses being much bigger. 
By computing the temperature dependence of the renormalization of the fourth order couplings, we find a width of the critical region of about $10^{-6} K$, in good agreement with the Ginzburg criterion.  
The resulting phase diagram is shown in figure  \ref{abb:PD1}. 
 In a pressure regime of $0.26<\delta<0.46$ a beak of A-phase between the symmetric and the B-phase is stabilized by the fluctuations of the order parameter. In the middle of this pressure interval the temperature width of this  wedge is about  $10^{-6} K$. 

\begin{figure}
\centering\epsfig{file=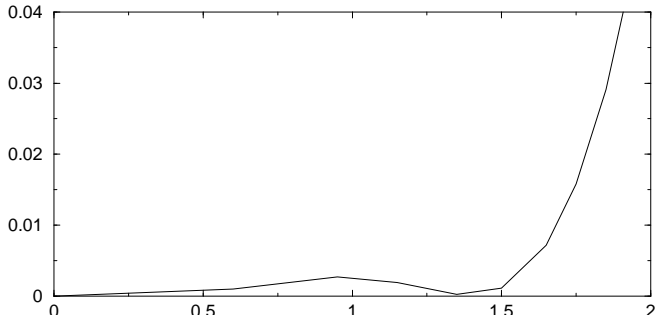,angle=-90,clip=,width=\linewidth}
\caption{Potential $U_{k_f}$ in planar direction at the phase transition. We show the dependence on $\xi =  \sqrt{\frac{\rho}{3}} $ }
\label{abb:potential}
\end{figure}

Towards the upper end it rises, however, to  $10^{-5} K$ at $\delta=0.41$. All transitions are of the first order. The potentials at the transitions look qualitatively like figure \ref{abb:potential}.

\begin{figure}
\centering\epsfig{file=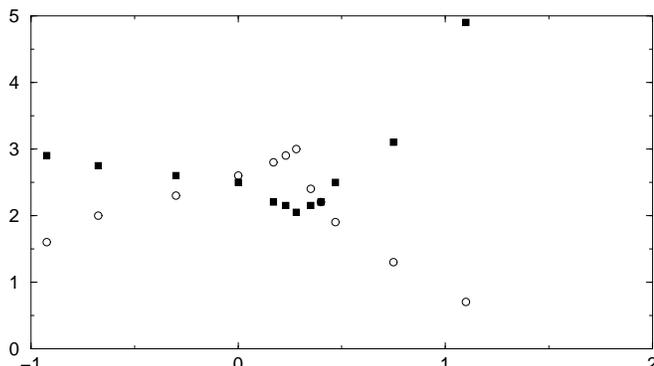,angle=-90,clip=,width=\linewidth}
\caption{Pressure dependence of the correlation length at the phase transition (in $\mu m$, circles) and the discontinuity in the NMR-frequency shift (in $kHz$, squares)}
\label{abb:m_von_T}
\end{figure}
 Figure \ref{abb:m_von_T} shows the pressure dependence of the correlation length, given by the inverse of the renormalized mass at the potential minimum, and the  discontinuity of the NMR-frequency shift at the phase transition.  The latter quantity is related to the jump in the order parameter squared $\triangle \rho$ by 

\begin{equation}
 \triangle \nu^2 \approx 6 c \cdot 10^{10}\;\; Hz  \cdot \triangle \rho
\end{equation}
which follows from the mean-field temperature dependence of $\triangle \nu$ given in \cite{Love}.
We observe a substantial increase of the correlation length as we approach the triple point $\delta=0.26$. This observation nourishes first hope for a fixed point in the vicinity of this point in parameter space which would give rise to universal behaviour of the system. Accurate predictions could then be made despite the uncertainty of our knowledge of the microscopic theory.
 Also an examination of the flow of the couplings at the origin looks promising for the triple point. We observe strong renormalization of the fourth order couplings $b_i$. For example, $\frac{b_{4_{\Lambda}}}{b_{4_0}} \approx 8$, whereas analogous ratios away from the triple point, say, at $\delta=0$,  grow no larger than $1.5$ at the phase transition .

\begin{figure}
\centering\epsfig{file=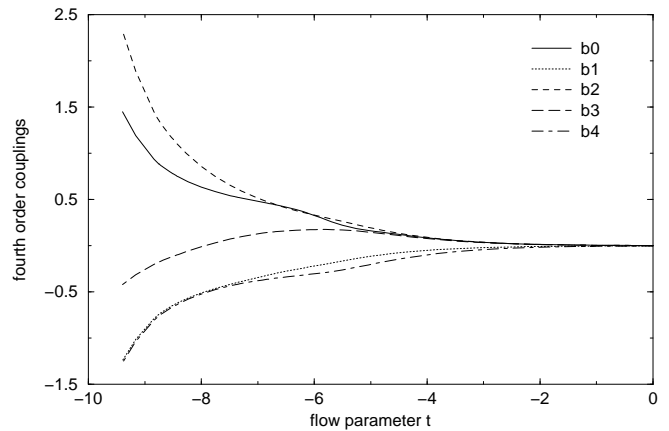,angle=-90,clip=,width=\linewidth}
\caption{Flow of the dimensionless  renormalized fourth order couplings near the triple point ($c=0.001$, no spin-orbit coupling) }
\label{abb:bi}
\end{figure}
Figure \ref{abb:bi} shows the dimensionless renormalized couplings $b_{i_r}$. We cannot find a true fixed point behaviour. However, ratios of couplings seem to tend to constant values, such as $\frac{b_{1_r}}{b_{4_r}} =1$, corresponding to  a partial fixed point.

\begin{figure}
\centering\epsfig{file=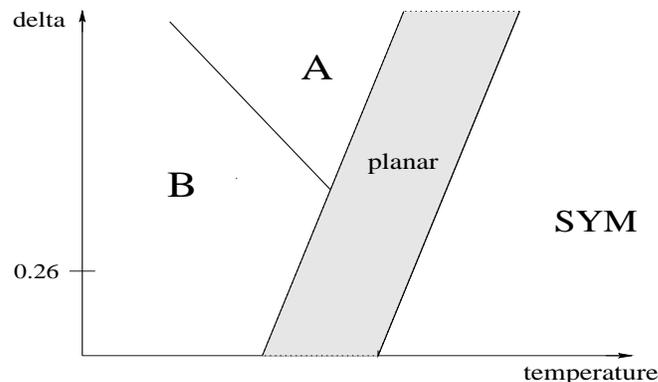,clip=,height=5cm,width=\linewidth}
\caption{Phase diagram of the effective theory for liquid $^3 He$ in the  parameter range suggested by the paramagnon model. The temperature width of the planar phase is about $10^{-9} K$.}
\label{abb:PD2}
\end{figure}
The qualitative features for smaller couplings $c=0.001$ look similar, with correspondingly smaller $\triangle T=10^{-8}K$. In this case, however, the dipole term should be included ($a_{\Lambda}=2\cdot 10^{-6}$) which we do next. Figure \ref{abb:PD2} shows the phase diagram with first order transition lines. 
 The considerations in \cite{Love} are confirmed by our calculation. The width of the stripe of planar phase stabilized by the dipole interaction is  about $2 \cdot 10^{-9} K$ as it was expected from mean-field considerations. We have not inspected the phase structure much further than $10^{-9} K$ away from the transition line. We expect the same features we found above: stabilization of a wedge of A phase with a temperature width of about $10^{-8} K$ by the fluctuations.

 Tuning the temperature to the transition from the symmetric to the planar phase we find a potential $U_{k_f}$ (figure \ref{abb:potential}) which shows clearly that the transition is of first order. (Fluctuations with momenta smaller than $k_f$ make the potential convex without much influence on the equation of state \cite{report}.) From figure \ref{abb:potential} we can easily infer the order parameter jump:
$ \triangle \rho \approx 8 \cdot 10^{-5}$.
This turns out to be almost independent of the pressure in the regime examined by us. It corresponds to a discontinuity in the  NMR-frequency shift at the transition of about $50\; Hz$. It was suggested in \cite{Love} that this should be experimentally observable.

Let us finally ask how experiments may constrain the size of the couplings. We have seen that large couplings lead to a widening of the critical region $\triangle T$. 
 It should be possible to exploit this fact by measuring the width of the critical region. Varying the temperature at a pressure right under the triple point,  one should be able to observe two distinct phase transitions . The width of the intermediate phase relates to the coupling strength via  the Ginzburg criterion.   By the described experiment one could also measure the pressure  corresponding to the values $\delta=0.26$ and $\delta=0.46$. An approximate form of the relation $p(\delta) $  could be obtained then by linearization.  Strictly speaking, these arguments hold only if the dipole term can be neglected. However, the temperature range for the planar phase, $10^{-9} K$, is  so narrow, that it would probably not even show up in experiments with realistic temperature resolution. 

\begin{figure}
\centering\epsfig{file=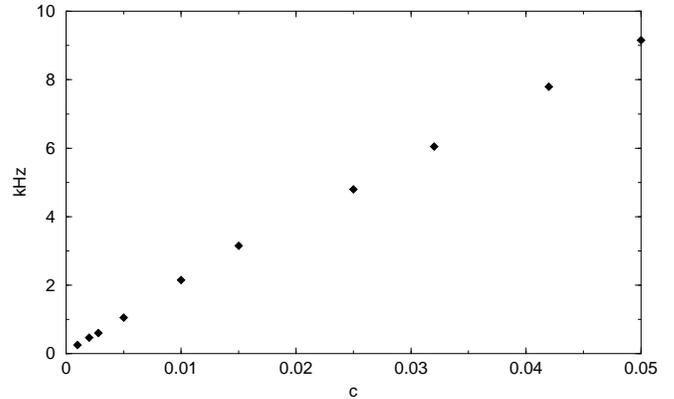,angle=-90,clip=,width=\linewidth}
\caption{Dependence of the discontinuity of the frequency shift (in $kHz$) on the coupling strength $c$ }
\label{abb:dnu_von_c}
\end{figure}

Also the order parameter jump depends strongly on the coupling strength. With the larger couplings $c=0.01$ we find $\triangle \rho \approx 0.01$ for the jump into the B-phase at the triple point. This leads  to a discontinuity in the frequency shift of about $2000 Hz$ as opposed to e.g. $370 Hz$ for $c=0.002$. We think, that  it would be even easier to infer the size of the critical region from measuring this order parameter jump. At least, this way it should be possible to get upper bounds for the coupling strength that are rather close to the range discussed in this article.  To facilitate the interpretation of such NMR-experiments, we calculated the jump of $\triangle \nu$  for various other values of the coupling strength c at $\delta=0.15$.
We show the result in figure \ref{abb:dnu_von_c}.

In conclusion, our quantitative description of the phase diagram of $^3He$ should permit to extract reliable information about the system's microphysical interactions from future experiments.

\end{multicols}


\begin{thebibliography}{00}

\bibitem[*]{mailkind}
e--mail: M.Kindermann@ThPhys.Uni-Heidelberg.DE

\bibitem[\dagger]{mailwett}
e--mail: C.Wetterich@ThPhys.Uni-Heidelberg.DE

\bibitem{Osheroff}
D.~D. Osheroff, R.~C. Richardson, and D.~M. Lee, Phys. Rev. Lett. {\bf 28}, 885 (1972)    .

\bibitem{Love}
D.~R.~T. Jones, A. Love, and M.~A. Moore, J. Phys C {\bf 9}, 743 (1976)   .

\bibitem{Sokolov79}
A.~I. Sokolov, JETP Letters {\bf 29},  590 (1979)  .


\bibitem{Wheatley}
Paulson and Wheatley, Phys. Rev. Lett. {\bf 41}, 497 (1978)   .


\bibitem{Sokolov83}
A.~I. Sokolov, Sov. Phys. JETP {\bf 57}, 798 (1983)   .

\bibitem{Wetterich}
C. Wetterich, Phys. Lett. B {\bf 301}, 90 (1993)    .

\bibitem{Bailin}
B. Bailin, A. Love, and M.~A. Moore, J. Phys C {\bf 10}, 1159 (1977)   .


\bibitem{Tetradis}
N. Tetradis and C. Wetterich, Nucl.Phys. B {\bf 422}, 541 (1994)   .

\bibitem{Vollhardt}
D. Vollhardt and P. W\"olfle, {\em The superfluid phases of Helium 3} (Taylor
  and Francis, London, 1990).


\bibitem{report} 
J. Berges, N. Tetradis, C. Wetterich,  hep-ph/0005122 (2000)  .
\end{thebibliography}
\end{document}